\documentclass[runningheads]{llncs}
\usepackage[T1]{fontenc}
\usepackage{graphicx}
\usepackage{amsmath,amssymb,amsfonts}%

\begin{document}

\title{ScentEcho: Exploring Adsorbent Materials for Accurate Odor Collection and 
Playback}
\titlerunning{ ScentEcho: Adsorbent Materials for Odor Collection and Playback}
%
\author{Chih-Hung Lee\inst{1,2}\orcidID{0009-0007-5949-389X} \and
Yuchi Sun\inst{1,2}\orcidID{0009-0000-8765-5541} \and
Rui Zhang\inst{2}\orcidID{0009-0007-1953-0844} \and Suhang Wei\inst{1}\orcidID{0009-0000-6860-0166} \and Qi Lu\inst{2}\orcidID{0000-0003-4507-1382}}
\authorrunning{Lee et al.}
%
\institute{Academy of Arts \& Design, Tsinghua University, Beijing, China \and
The Future Laboratory, Tsinghua University, Beijing, China
\email{luq@mail.tsinghua.edu.cn}}
\maketitle              

\begin{abstract}
Delivering odors that feel realistic and recognizable remains a core challenge for olfactory interaction systems, particularly in applications that demand precise scent delivery. A key limitation lies in the difficulty of capturing, preserving, and playing back real-world scent sources in a reliable and scalable manner. This study explores the potential of adsorbent materials for supporting realistic scent playback. We present ScentEcho, a portable system that enables modular scent collection and release. Through user evaluations, we identify which adsorbent materials tend to perform better for specific odors, and observe that perceived intensity strongly influences similarity ratings. In addition, odor recognition follows a graded pattern, with users moving from broad category identification to more specific source recognition as similarity increases. These findings offer practical insights for designing olfactory interfaces that are both expressive and perceptually aligned with user expectations.

\end{abstract}

\section{INTRODUCTION}\label{sec:introduction}

In daily life, vision and hearing account for about 90\% of our sensory input \cite{iseki2022study}. While the sense of smell occupies a smaller proportion, it has a unique ability to evoke emotions and memories \cite{krusemark2013sense}. Leveraging this characteristic, HCI (Human-Computer Interaction) research explores how integrating smell can create more immersive experiences and enhance user engagement \cite{amores2017essence,brooks2023smell,lu2023atmospheror,obrist2014opportunities}.

However, current systems often rely on pre-defined odors \cite{covington2018development,zhang2024odoragent}, such as preset fragrances like the smell of roses, which may not align with individual perceptions and memories. While some studies enable users to customize mixtures by adjusting proportions of base compounds \cite{fu2024aromablendz,prasetyawan2024odor}, these approaches remain constrained by a fixed set of source materials. As a result, they hinder the personalization and contextual adaptability of olfactory interaction, making it difficult for users to engage with scents that authentically reflect their experiences, surroundings, or emotional states.

To address these challenges, one potential direction is to allow users to obtain scents directly from their surroundings, giving them the ability to collect and preserve odors they find personally meaningful or interesting. Prior work by designers and companies has proposed various scent collection concepts \cite{mohan2017smellcamera,murins2021scentcamera,radcliffe2013madeleine}, but many remain exploratory or artistic in nature without validating the technical performance of materials or systems. Notably, the ``smell camera'' \cite{lu2020exploring}, presents a creative approach to scent capture and playback, yet issues such as bulkiness, limited reusability, and the inability to store or switch between multiple scents still hinder practical deployment. These gaps highlight the importance of systematically investigating materials and methods for achieving reliable and realistic odor capture in real-world contexts.

We therefore developed ScentEcho, a compact and portable device designed to support personalized scent collection and playback. Its modular structure and simple pneumatic system allow for the storage and release of multiple odors in a lightweight format. In this study, we employed ScentEcho to evaluate the performance of different adsorbent materials in capturing and reproducing commonly studied odors in HCI, such as lemon, coffee bean, mint, and lavender \cite{brianza2019light,fraune2022scent,garcia2008integrating,herz2004changing,madzharov2018impact}. Participants assessed the similarity and intensity of the replayed scents. In addition to evaluating material performance, our study also revealed a notable correlation between perceived intensity and odor similarity, as observed through participant ratings and interviews. This insight offers a deeper understanding of how users perceive olfactory realism and the factors that influence their interpretation of scent fidelity. 

\section{RELATED WORK}\label{sec:related-work}
\subsection{Scent Sources in Olfactory Interaction}\label{subsec:scent-sources}
Olfaction has gained significant attention in HCI, with research focusing on its integration with other sensory modalities to create immersive and emotionally engaging user experiences \cite{gao2024mul,lei2022diy,wang2024olfackit,yan2023scentcarving}. However, most existing studies rely on commercially available sources such as perfumes or synthetic fragrance compounds \cite{brooks2023smell,fei2023laseroma,fu2024aromablendz,lu2023atmospheror,you2022fragrance}. While these sources are convenient, their labels or descriptors often fail to accurately reflect olfactory perception. Prior studies have indicated that verbal scent descriptors do not always align with individual olfactory perception, suggesting that even familiar terms may be interpreted differently across users. While these findings emerge primarily from linguistic and cognitive research, they point to the challenge of relying on predefined scent labels in olfactory interface design \cite{kaeppler2013odor,iatropoulos2018language}.

Although not all studies explicitly emphasize realism or semantic alignment, these aspects are especially valuable when olfactory cues are used to evoke emotion, trigger memory, or support interpersonal communication. To better align with users' lived experiences and olfactory memories, systems should move beyond predefined commercial fragrances and support more diverse, personalized scent inputs that reflect real-life contexts and meaningful interactions.

\subsection{Scent Collection and Playback}\label{subsec:scent-collection}
To support more authentic olfactory interaction, researchers and designers have proposed various scent collection and playback systems. Traditional methods such as solvent extraction or hydrodistillation can chemically capture real-world odors with high accuracy \cite{azmir2013techniques,gallagher2008analyses,lucchesi2004solvent}, but they are typically time-consuming, equipment-intensive, and require large quantities of raw materials to produce even small amounts of extract. This limits their feasibility in interactive systems, particularly those aimed at supporting personal, context-aware, or real-time scent experiences.

In response, designers and companies have explored alternative approaches, proposing physical devices for capturing and re-releasing target odors \cite{mohan2017smellcamera,murins2021scentcamera,radcliffe2013madeleine}. While such designs introduce creative interaction concepts, many remain exploratory or artistic in nature and lack rigorous evaluation of technical performance. Among prior work, the ``smell camera'' \cite{lu2020exploring} offers an illustrative example of a novel HCI interface exploring user expectations for scent collection---such as duration, convenience, and intended use. While offering valuable insights, its practical deployment remains limited due to bulkiness, one-time usability, and the inability to store or switch between multiple scents. Crucially, most existing systems do not assess whether the released scent retains perceptual similarity to its source, leaving a fundamental gap in understanding the fidelity of olfactory reproduction.

\section{DESIGN AND IMPLEMENTATION }\label{sec:design}
\begin{figure}[htbp]
  \centering
  \includegraphics[width=0.8\linewidth]{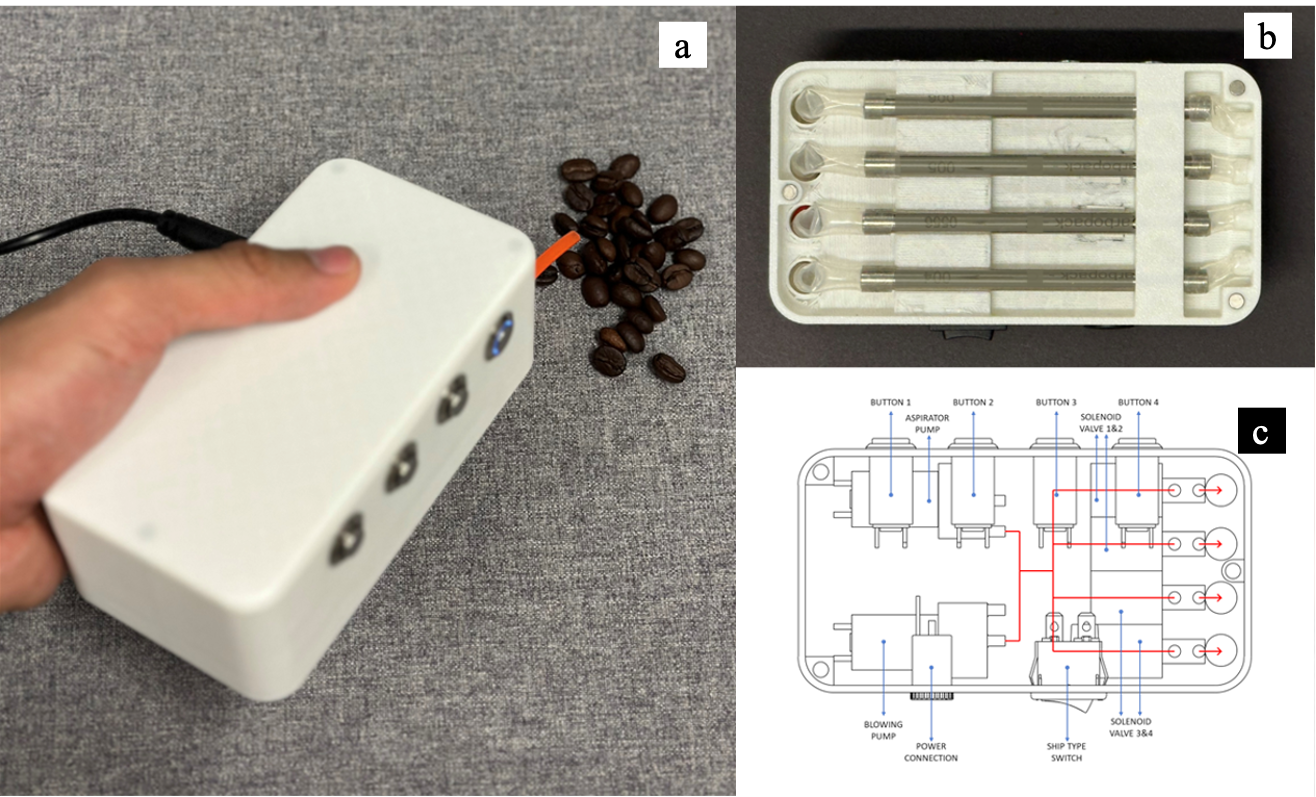} 
  \caption{: Illustrations of the Device Structure and Operation. a) Schematic of the odor collection process, b) Internal view of the upper compartment, c) Schematic of the upper compartment structure and pneumatic pathways.}
  \label{fig:scentecho}
\end{figure}

ScentEcho is composed of two separate compartments: the upper compartment, which houses the electronic components 
and serves as the control unit, and the lower compartment, which contains four independent odor storage units, forming 
the device's core functional component. Fig.~\ref{fig:scentecho} illustrates the overall structure and operation of ScentEcho. 

The upper compartment includes four self-locking buttons, a three-position toggle switch, and a power connection. It 
houses two air pumps (5V, 1.1 L/min) for odor collection and playback. To ensure pump longevity and odor delivery 
stability, a toggle switch ensures only one pump operates at a time. Four two-way solenoid valves (5V) are individually 
controlled by self-locking buttons. For ease of maintenance, the lower compartment features only the odor storage units 
and support structures. Magnets connect the upper and lower compartments, facilitating easy inspection and replacement. 
Fig.~\ref{fig:scentecho}(b) details the internal structure of the compartments.

The pneumatic system is depicted in Fig.~\ref{fig:scentecho}(c), with red lines indicating airflow. Air pumps provide the 
driving force, and solenoid valves control specific pathways. Two pumps connect to four solenoid valves via three-way 
and four-way connectors, and these valves are linked to the odor storage units in the lower compartment through air hoses. 
Users operate the device by pressing one or more self-locking buttons to activate the corresponding solenoid valves, 
followed by toggling the switch to start the aspirator or blowing pump for odor collection or playback. Backlit buttons 
provide visual feedback, indicating the active odor storage unit, as shown in Fig.~\ref{fig:scentecho}(a).

ScentEcho offers several key advantages: 1. Odor Collection and Playback, it can manage four different odors 
simultaneously, with reserved space in the lower compartment for additional storage units, enabling expansion; 2. Compact 
Design and Maintenance, the device's small size (127$\times$68$\times$47 mm), light weight (248 g), and simple structure make it easy 
to operate and maintain with one hand. 3. Simple Controls, no programming or integrated circuits are required, as 
functionality is managed by four buttons and a toggle switch, ensuring user-friendliness. 4. Portability and Power Supply, 
although it lacks a built-in battery, the device is designed to connect to a power bank via a simple adapter, with all 
components running on a 5V power supply.

After implementation, ScentEcho was employed to evaluate and select suitable adsorption materials for its odor storage 
units. The device's efficient and user-friendly design proved highly effective for testing the performance of different 
materials in capturing and replaying various odors.

\section{SCENT PLAYBACK PERFORMANCE EVALUATION}\label{sec:evaluation}
This study evaluates the performance of different adsorbent materials in capturing odorants, primarily through user experiments focusing on the similarity and intensity of reproduced odors. In addition to quantitative ratings, we conducted qualitative interviews to understand how people experience odor fidelity and form expectations about the reproduction of everyday scents.
\subsection{Experiment Setup}\label{subsec:experiment-setup}
This study evaluates the collection and playback performance of various adsorbent materials for odorants. The selected 
odorants, commonly used in HCI research, include lemon, coffee beans, mint, and lavender, as shown in Fig.~\ref{fig:Odor_and_Sniff}(a), all purchased 
from local markets. Coffee beans and lavender were stored at room temperature (25$^\circ$C), while fresh lemons and mint leaves 
were refrigerated at -4$^\circ$C. Before the experiments, refrigerated samples were brought to room temperature for adsorption. 
The specifications of adsorbent materials were listed in Table~\ref{tab:adsorbents}. The materials were packed into stainless steel tubes (outer 
diameter: 6 mm, length: 90 mm) with a particle size of 60-80 mesh and a loading amount of 200 mg. Before use, the 
adsorption tubes were purged at 250$^\circ$C for 30 minutes and cooled to room temperature.

All experiments were conducted under controlled conditions of 25$\pm$2$^\circ$C and 50$\pm$10\% relative humidity. Prior to the experiment, all participants signed an informed consent form, indicating their understanding of the experiment's procedure 
and their right to withdraw at any time.

\begin{table}
\centering
\caption{: Specifications of adsorbent materials used in this study.}\label{tab:adsorbents}
\begin{tabular}{|l|l|l|}
\hline 
\textbf{Code} & \textbf{Type} & \textbf{Supplier's Note} \\
\hline 
A & Tenax TA & C7--C26\textsuperscript{a} \\
B & Carbopack X & C3--C6 \\
C & AC & -- \\
D & T-CXmix & Tenax TA + Carbopack X \\
\hline 
\end{tabular}

\vspace{0.5em} 
\noindent\textsuperscript{a} Carbon number range of substances suitable for adsorption.
\end{table}

\subsection{User Evaluation Procedure}\label{subsec:user-procedure}
Both odor collection and playback were conducted using the ScentEcho device. 
During odor collection, the target odors (lemon, coffee beans, mint, and lavender) were adsorbed onto the selected 
adsorbent materials (A-D). The collection time was set to 1 minute, with minor adjustments based on a study by Lu et al. 
\cite{lu2020exploring} investigating acceptable durations for odor collection. Then, participants were exposed to the reproduced odors (playback) and asked to rate their similarity and intensity relative to the target odors using a Likert scale \cite{stoklasa2017fuzzified,torriani2024perceived}, where the target odor was defined as 7 points (1 = completely dissimilar or no odor; 7 = identical to the target).  To evaluate the relationship between odor similarity and intensity ratings, the Pearson Correlation Coefficient $\gamma$ was used to assess the relationship between odor similarity and intensity, calculated as Equation 1, where, $X$ and $Y$ represent similarity and intensity ratings, while $\bar{X}$ and $\bar{Y}$ are their means.

\begin{equation}
\gamma = \frac{\sum (X - \bar{X})(Y - \bar{Y})}{\sqrt{\sum (X - \bar{X})^2 \sum (Y - \bar{Y})^2}}
\label{eq:pearson}
\end{equation}
\vspace{0.5em}

\begin{figure}[htbp]
  \centering
  \includegraphics[width=0.75\linewidth]{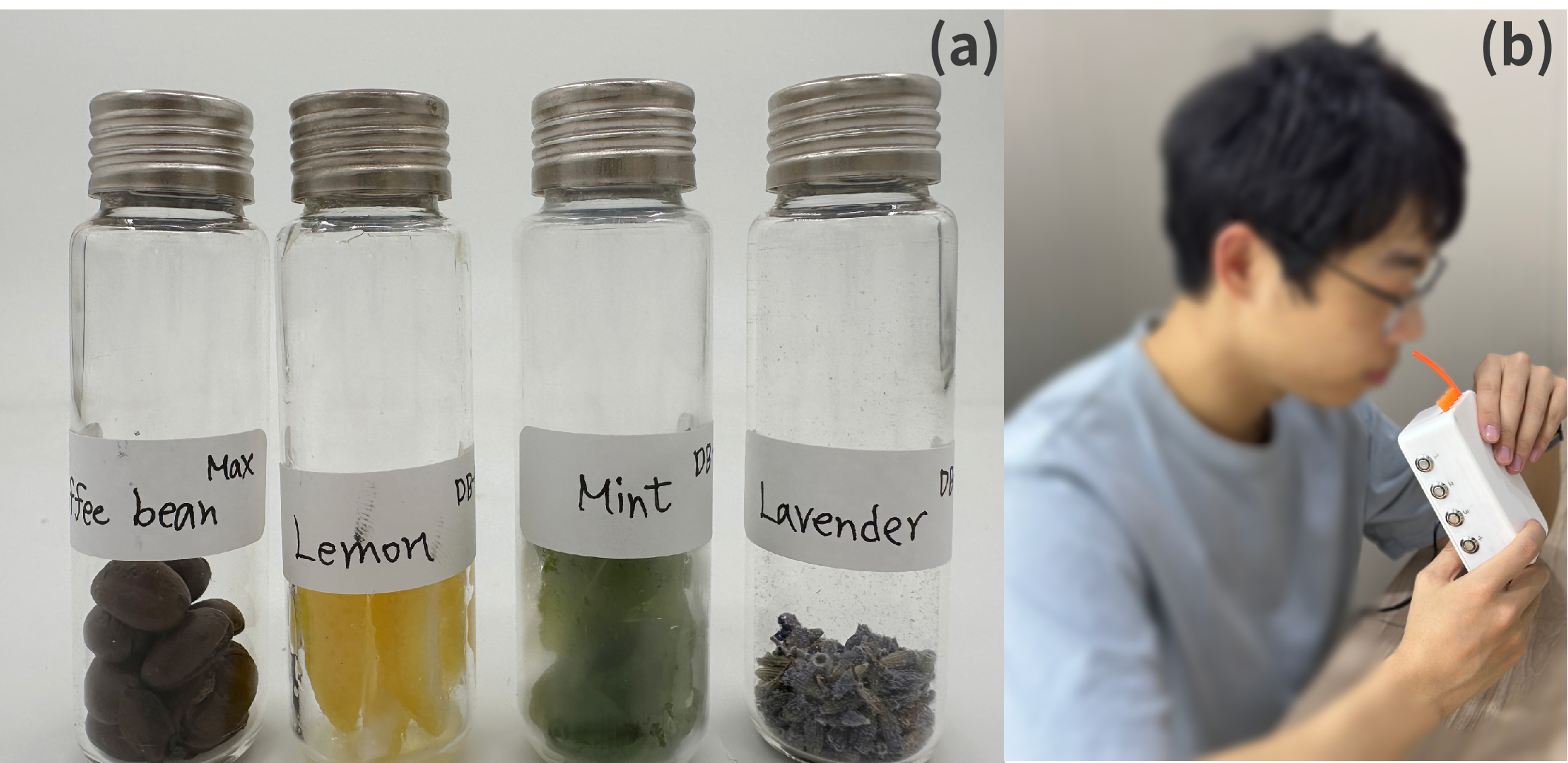}  
  \caption{: Evaluation and analysis. a) Odor materials used in the experiment, from left to right: coffee beans, fresh lemon, mint leaves, and dried lavender, b) A participant using ScentEcho to evaluate odor similarity and intensity. }
  \label{fig:Odor_and_Sniff}
\end{figure}

Participants were instructed to rate the reproduced odors within 10 seconds and were allowed to re-smell the target odor as needed---see Fig.~\ref{fig:Odor_and_Sniff}(b)
. The rating form (Appendix A) collected personal information and scoring data, as well as responses to 
questions such as: ``Do you regularly use perfumes or fragrance-related products?'' and ``Do you consider yourself to have 
a keen sense of smell?'' Additionally, exploratory questions were posed during the experiment to further investigate 
participants' perceptions. These included: ``Can you independently evaluate the similarity and intensity of an odor?'' and 
``At what similarity score do you feel confident in identifying the odor source without needing to see the actual object?'' 
The experiment consisted of two main stages. Odor preparation, including collection and playback setup, required 
approximately 40 minutes. And participant evaluations took an additional 30 minutes. To minimize fatigue and ensure 
accuracy, a 3-minute rest period was provided between exposure to different target odors.

\section{RESULTS}\label{sec:results}
This study recruited 12 participants (6 males and 6 females) aged between 22 and 49 years, with an average age of 31.2 years (standard deviation: 7.3). Regarding the question, ``Do you regularly use perfumes or fragrance-related products?'', 2 participants reported frequent use, 9 reported occasional use, and 1 reported no use. For self-assessed olfactory sensitivity, 1 participant strongly agreed they were sensitive, 5 agreed, 4 were neutral, and 2 disagreed. Overall, most participants had experience using perfumes or fragrances, but their self-perception of olfactory sensitivity varied widely. These background characteristics help contextualize participants' olfactory evaluations in the subsequent analyses.

\subsection{Ratings of Odor Playback Fidelity}\label{subsec:ratings}
\begin{figure}[htbp]
  \centering
  \includegraphics[width=1.00\linewidth]{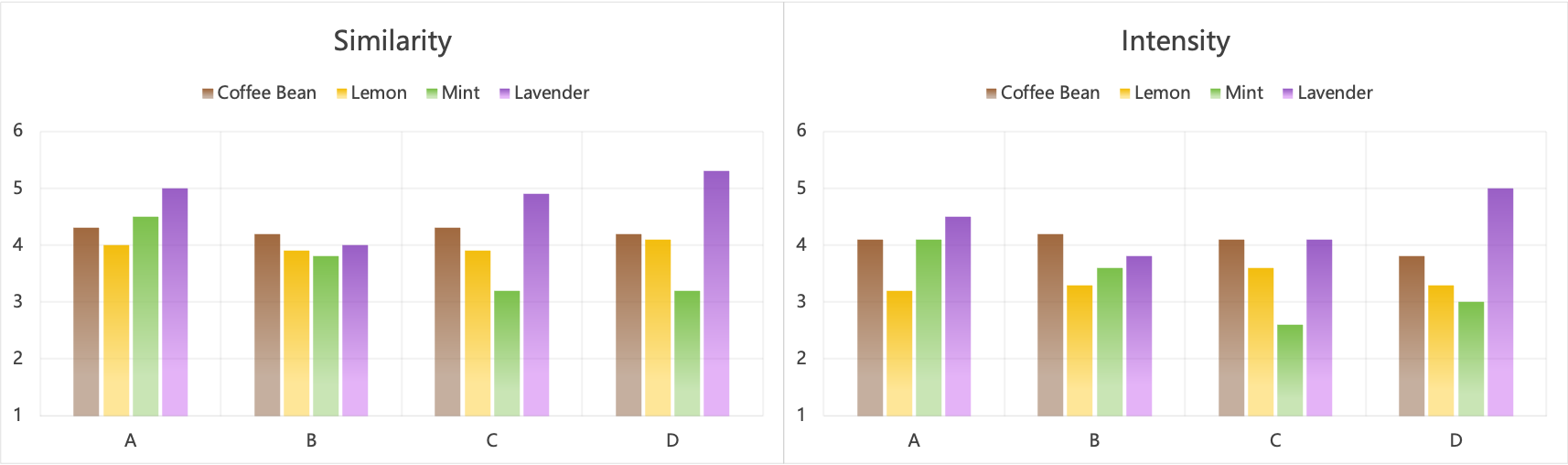}  
  \caption{: The score of odor similarity (left) and intensity (right). }
  \label{fig:Score}
\end{figure}

 The similarity and intensity rating results are shown in Fig.~\ref{fig:Score}. For similarity, different adsorbent materials showed varying performance across odor types. For example, material A performed well with mint, receiving an average similarity score of 4.5, while material D achieved a relatively high score of 5.3 for lavender. The similarity scores for coffee and lemon odors showed relatively small differences across the four adsorbent materials (coffee: 4.2--4.3; lemon: 3.9--4.1). In terms of intensity, material D stood out for lavender (5), whereas materials A, B, and C all achieved relatively high intensity scores for coffee (above 4). Overall, the findings suggest that different adsorbent materials are better suited for different types of odors, with their effectiveness varying across both similarity and intensity dimensions.

\subsection{User Perceptions of Odor Similarity and Intensity}\label{subsec:perceptions}
\begin{figure}[htbp]
  \centering
  \includegraphics[width=0.6\linewidth]{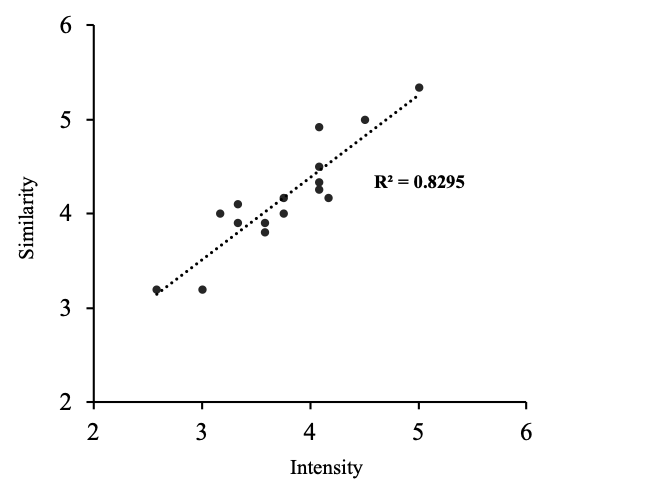}  
  \caption{: Correlation between scores of odor similarity and intensity (across all odors and participants). }
  \label{fig:XY}
\end{figure}

Analysis of user ratings revealed a strong positive correlation between perceived similarity and intensity, with a Pearson correlation coefficient of 0.8295 (Fig.~\ref{fig:XY}). This indicates that similarity ratings tended to increase with perceived odor strength.

During the experiment, participants were asked whether they could evaluate similarity and intensity independently. Among the 12 participants, 7 reported difficulty in distinguishing the two dimensions, particularly when the reproduced scent was weak. The remaining 5 participants indicated that they could evaluate them separately when intensity was sufficiently strong.

Participants were also asked how similarity ratings related to odor recognition. At similarity scores of 3--4, users tended to associate the scent with general categories (e.g., mint with toothpaste or mint candy). At scores of 5--6, they were able to identify more specific odor sources, such as ``fresh mint leaves.''

In this study, several adsorbent materials achieved playback that received user-rated similarity scores above 5 for specific target odors. This indicates that the system was capable of reproducing scents with a high degree of perceptual fidelity, where the released odors closely matched the original odor sources as experienced by users.

\section{ DISCUSSION }\label{sec:discussion}
This section examines how odor similarity and intensity influence perception, and how different adsorbent materials impact recognition. We discuss key design challenges in olfactory interaction based on user feedback and perceptual evaluation results.

\subsection{The Interdependence of Similarity and Intensity in Odor Perception}\label{subsec:intensity-similarity}
A strong correlation was found between perceived intensity and similarity (r = 0.8295), suggesting that users often conflate these two dimensions when evaluating odors. During the experiment, 7 out of 12 participants reported being unable to distinguish similarity from intensity, especially when the reproduced scent was weak. However, 5 participants noted that once the scent reached a certain intensity threshold, they could evaluate similarity more independently. This reveals a key challenge in olfactory perception---where reduced intensity can obscure odor characteristics and impair recognition.

Interestingly, when asked about auditory perception as a comparison, most participants (8 out of 12) indicated they could distinguish between timbre (analogous to odor similarity) and volume (analogous to odor intensity). This contrast suggests that olfactory perception may involve more entangled perceptual dimensions than auditory ones, potentially due to its chemical basis and less practiced use in discrimination tasks.

These findings imply that odor playback systems must ensure not only that the scent composition reflects the original odor profile, but also that the intensity is sufficient for users to perceive the intended similarity. Therefore, intensity should be considered a foundational design parameter, especially in use cases that demand precise recognition.

\subsection{Hierarchies of Odor Recognition and Design Implications}\label{subsec:recognition-hierarchy}
User feedback also revealed a perceptual hierarchy in odor identification. At similarity scores of 3--4, participants could identify general categories (e.g., associating mint with toothpaste or mint candy), while at scores of 5--6, they could pinpoint specific odor sources such as ``fresh mint leaves.'' This observation suggests that olfactory recognition occurs along a graded spectrum, and system designers may leverage this knowledge to define fidelity goals based on context.

For instance, in ambient experience design, evoking general odor categories may suffice to create emotional or atmospheric effects. In contrast, applications requiring precision (e.g., culinary training or fragrance reproduction) may demand more accurate scent reconstruction. Thus, the level of perceptual resolution achievable through a given system should be aligned with its intended purpose.

\subsection{Practical Considerations and Future Directions}\label{subsec:future-directions}
This study identified several practical considerations that may inform future refinement of the ScentEcho system. First, the sample size was limited to 12 participants aged 23--35, whose olfactory sensitivity may not reflect broader user populations. Second, the intensity of the reproduced odors was generally weaker than the original, as reported by multiple participants. While longer adsorption times might enhance scent strength, this would also increase preparation time and reduce responsiveness in real-time applications. Third, some participants noticed faint metallic smells when testing weaker odor samples, likely due to the use of stainless steel adsorption tubes. Additionally, the current design activates the air pump even when all solenoid valves are off, potentially affecting the pump's lifespan and compromising airtightness.

To address these issues, several improvements are planned. These include refining the circuit to ensure the air pump activates only when an airflow pathway is open, and incorporating a heating element to enhance control over release intensity. Additionally, inert alternatives such as fluoropolymers will be investigated to minimize material-related interference. Future experiments will also include a wider range of scent compounds, from light to heavy odors, to better reflect the diversity of smells encountered in real-world settings. To improve odor reproduction fidelity, we plan to evaluate modified or composite adsorbent materials and consider integrating electronic nose systems for continuous monitoring.

Despite these limitations, ScentEcho demonstrates strong potential as a modular scent playback platform for HCI. Its flexible configuration and compatibility with multiple adsorbent materials make it a promising tool for applications in personalized scent experiences, scent-triggered memory recall, multisensory storytelling, and sensory training.

\section{ CONCLUSION }\label{sec:conclusion}

This study presents ScentEcho, a portable system for personalized scent collection and playback. By integrating modular hardware and adsorbent material evaluation, it enables realistic odor reproduction beyond predefined fragrances. User studies revealed how scent similarity and intensity interact in perception, underscoring the importance of human-centered evaluation in material selection. ScentEcho bridges a key gap in HCI by supporting more contextual and emotionally resonant olfactory interactions. Although the current prototype supports four odor channels, its modular structure is designed to accommodate future expansion through additional units and improved component integration.  Its flexible design opens new possibilities in applications such as targeted scent delivery, memory recall, and multisensory storytelling.

\begin{credits}
\subsubsection{\ackname} 
This work was supported by the Foundation of the National Key Laboratory of Human Factors Engineering (Grant No. HFNKL2023J08) and the National Natural Science Foundation of China (Grant Nos. 62441219 and 62172252).

\subsubsection{\discintname}
The authors have no competing interests to declare that are
relevant to the content of this article.
\end{credits}

\appendix
\section{Odor Similarity and Intensity Rating Form (For User Experimentation)}\label{sec:rating-form}

\textbf{I. Personal Information}

\vspace{0.3em}

Name: \underline{\hspace{3cm}} \hspace{2em} Gender: \underline{\hspace{2cm}} \hspace{2em} Age: \underline{\hspace{2cm}}

\vspace{1em}

1. Do you usually have the habit of using perfumes or fragrance-related products?\\
\hspace*{1em}$\Box$ Never \quad $\Box$ Occasionally \quad $\Box$ Sometimes \quad $\Box$ Often \quad $\Box$ Daily

\vspace{0.5em}

2. Do you consider yourself a person with a keen sense of smell?\\
\hspace*{1em}$\Box$ Strongly Disagree \quad $\Box$ Disagree \quad $\Box$ Neutral \quad $\Box$ Agree \quad $\Box$ Strongly Agree

\vspace{1em}

\textbf{II. Odor Similarity and Intensity Evaluation}

Instructions for Scoring: The similarity and intensity of the standard sample are defined as 7 points.  
Please use this as a reference to evaluate the scores for samples A--H.

(1 = Completely dissimilar or no odor at all; 7 = Identical to the standard sample)

\vspace{1em}

\begin{table}
\centering
\begin{tabular}{|c|cc|cc|}
\hline
\textbf{Code} & \multicolumn{2}{c|}{\textbf{Target Odor 1}} & \multicolumn{2}{c|}{\textbf{Target Odor 2}} \\
              & Similarity (1--7) & Intensity (1--7) & Similarity (1--7) & Intensity (1--7) \\
\hline
A & & & & \\
B & & & & \\
C & & & & \\
D & & & & \\
\hline
\end{tabular}
\end{table}

\vspace{1em}

\begin{table}[h]
\centering
\begin{tabular}{|c|cc|cc|}
\hline
\textbf{Code} & \multicolumn{2}{c|}{\textbf{Target Odor 3}} & \multicolumn{2}{c|}{\textbf{Target Odor 4}} \\
              & Similarity (1--7) & Intensity (1--7) & Similarity (1--7) & Intensity (1--7) \\
\hline
A & & & & \\
B & & & & \\
C & & & & \\
D & & & & \\
\hline
\end{tabular}
\end{table}




%
%
\bibliographystyle{splncs04_change} 
\bibliography{sn-bibliography.bib} 

\end{document}